# Nonlocality test of energy-time entanglement via nonlocal dispersion cancellation with nonlocal detection

Baihong Li[1,2,3], Feiyan Hou[1,2*], Runai Quan[1,2], Ruifang Dong[1,2‡], Lixing You[4,5†], Hao Li[4,5], Xiao Xiang[1,2], Tao Liu[1,2], and Shougang Zhang[1,2]

[1] *Key Laboratory of Time and Frequency Primary Standards, National Time Service Center, Chinese Academy of Sciences, Xi'an 710600, China*
[2] *School of Astronomy and Space Science, University of Chinese Academy of Sciences, Beijing, 100049, China*
[3] *College of Sciences, Xi'an University of Science and Technology, Xi'an, 710054, China*
[4] *State Key Laboratory of Functional Materials for Informatics, Shanghai Institute of Microsystem and Information Technology, Chinese Academy of Sciences, Shanghai 200050, China*
[5] *Center for Excellence in Superconducting Electronics, Chinese Academy of Sciences, Shanghai, 200050, China*
(Received  ;    published  ;)

Energy-time entangled biphoton source plays a great role in quantum communication, quantum metrology and quantum cryptography due to its strong temporal correlation and capability of nonlocal dispersion cancellation. As a quantum effect, nonlocal dispersion cancellation is further proposed as an alternative way for nonlocality test of continuous variable entanglement via the violation of Bell-like inequality proposed by Wasak et al. [Phys. Rev. A, **82**, 052120 (2010)]. However, to date there is no experimental report either on the inequality violation or on a nonlocal detection with single-photon detectors at long-distance transmission channel, which is key for a true nonlocality test. In this paper, we report an experimental realization of a violation of the inequality after 62km optical fiber transmission at telecom wavelength with a nonlocal detection based on event timers and cross-correlation algorithm, which indicates a successful nonlocal test of energy-time entanglement. This work provides a new feasibility for the strict test of the nonlocality for continuous variables in both long-distance communication fiber channel and free space.

## I. INTRODUCTION

Nonlocality is considered as a central feature of quantum entanglement, which cannot be explained by classical or any local hidden-variable theories [1-2]. Such nonlocality can be tested via a violation of Bell inequality in two different scenarios. The first scenario is to use discrete variables such as polarizations of the optical field. Due to its robustness to channel losses and noise, polarization-entanglement has been applied for more than 36 years to the test of Bell inequality violation and such violation has been recently realized over free-space links of thousands of kilometers by satellite [3-4]. The second scenario is to use continuous variables such as amplitude and phase quadratures of the optical field, which recently has been demonstrated in Ref. [5]. However, such quadratures will experience unavoidably large loss and decoherence in the long-distance transmission, making it difficult to verify nonlocality over long distances.

The energy-time entangled photons (biphotons), which has both strong temporal correlation and frequency anti-correlation, is intrinsically robust against the loss and decoherence when propagating through long-distance fiber links [6], it thus has found great applications in areas such as fiber-based quantum communication, quantum metrology [7-9] and quantum cryptography [10-11]. The nonlocality test of energy-time entanglement has been proposed by Franson with Franson-type interferometer [12] and nonlocal dispersion cancellation (NDC) [13]. The NDC is a nonlocal quantum effect, in which a pair of energy-time biphotons propagates through two distant dispersive media with equal and opposite dispersion, the dispersion experienced by one photon can be canceled nonlocally by the other. Benefited from the NDC effect, the energy-time biphotons will still remain tightly correlated in time after dispersion propagation. In contrast to another dispersion cancellation scenario [14, 15], in which the time measurement is based on the *local* Hong-Ou-Mandel [16] interference and has been proven to have classical analogues [17-21], the NDC effect is fundamentally independent of the separation between the two photons and provides a further example of the nonlocal nature of the quantum theory [22-25]. To witness the presence of entanglement in NDC, a Bell-like inequality has been further proposed by Wasak et al. [25] for bounding the strength of temporal correlations in a pair of light beams after propagating through dispersive media with equal and opposite dispersion. The violation of the inequality can thus be used for identifying the nonlocal feature of energy-time entanglement.

*houfeiyan@ntsc.ac.cn
‡dongruifang@ntsc.ac.cn
†lxyou@mail.sim.ac.cn

Since it was first proposed, the NDC effect has been demonstrated successfully at nanosecond-level resolution with local observation such as time-correlated single-photon counting (TCSPC; e.g., PicoHarp 300) [26-28]. A femtosecond-level NDC [29-30] has been achieved by up-conversion of the photon pairs [31-32], but is intrinsically local and can be mimicked using classical laser light [20], which limits its further application in a genuine quantum nonlocality test. Therefore, a *nonlocal detection* is particularly desired for a real test.

Recently, MacLean et al. [33] has already showed the violation of Wasak's inequality using optically gated single-photon detection technology in a nonlocal way, which enables to directly observe the NDC on femtosecond time scales. However, due to relatively large timing jitter of the single-photon detectors, the experimental violation of the inequality with distinct single-photon detectors at long-distance transmission channel has not been realized so far to our knowledge.

In this paper, we report an experiment that violates the inequality for the NDC with a nonlocal temporal correlation identification of biphotons. The energy-time entangled signal and idler photons at telecom wavelength are generated via the spontaneous parametric down conversion (SPDC) from a CW pumped Type-II PPKTP crystal [34]. Then the signal photons are transmitted through a 62-km-long single-mode fiber (SMF) while the idler photons are passed through a 7.47-km-long dispersion compensation fiber (DCF). To reduce the timing jitter of the detectors, self-developed superconducting nanowire single photon detectors (SNSPDs) [35] are applied, and the registered arrival times of the detected signal and idler photons are recorded independently by two separate event timers (ETs). By applying cross-correlation algorithm to these time sequences, the temporal correlation between the signal and idlers are then nonlocally identified [36]. In this way, we successfully demonstrate a NDC experiment which violates Wasak's inequality by about 14 standard deviations. This violation provides an experimental proof for the quantum nonlocality feature of continuous variable entanglement and can be effectively extended to long-distance transmission channel for further loophole-free test [37-40].

## II. THEORY

In this section, a review on the theory of the NDC and Wasak's inequality is briefly given. For a CW pumped type-II SPDC, when biphotons travelled through two dispersive media with dispersion coefficients of $k_s''$, $k_i''$ and lengths of $l_1$, $l_2$, the joint detection probability of the two detectors is proportional to the second-order Glauber correlation function [27]

$$G^{(2)}(t_1 - t_2) \propto e^{-\frac{[(t_1-t_2)+\bar{\tau}]^2}{2\sigma^2}}, \quad (1)$$

where $\sigma = \sqrt{\gamma D^2 L^2 + [(k_s''l_1 + k_i''l_2)/2]^2 / \gamma D^2 L^2}$ and $\bar{\tau} = k_s'l_1 - k_i'l_2$ denote the width of $G^{(2)}$ function and the overall time delay between signal and idler photons after propagating through the dispersion media, respectively. $L$ is the crystal length, $D$ is the inverse group velocity difference between signal and idler photons in the SPDC crystal, respectively. $\gamma = 0.04822$ is chosen for Gaussian approximation of the phase matching function. The full width at half maximum (FWHM) of $G^{(2)}$ function is then given as $\Delta t = 2\sqrt{2\ln 2}\sigma$.

If $k_s''$ and $k_i''$ have opposite signs, $|k_s''l_1 + k_i''l_2| \to 0$ can be realized by adjusting the lengths of $l_1$ and $l_2$. In the case of ideal NDC, $k_s''l_1 + k_i''l_2 = 0$ and the FWHM of $G^{(2)}$ function remains as $\Delta t = 2\sqrt{2\ln 2\gamma}DL$. So the effect of NDC can be observed by measuring the width of $G^{(2)}$.

On the other hand, the quantum nonlocality feature of the NDC can be verified by violation of a Bell-like inequality. With regard to this, Wasak et al. [25] deduced the minimum broadening of temporal correlations between two classical light beams during propagation through dispersive media with equal and opposite dispersion, which can be expressed as an inequality

$$\langle(\Delta\tau')^2\rangle \geq \langle(\Delta\tau)^2\rangle + \frac{(2\beta l)^2}{\langle(\Delta\tau)^2\rangle}, \quad (2)$$

where $\langle(\Delta\tau)^2\rangle$ and $\langle(\Delta\tau')^2\rangle$ are the time-difference variance before and after dispersive propagation, respectively. In our case, $2\beta l = |k_s''l_1| = |k_i''l_2|$. According to Ref. [25], a violation of this inequality is an unambiguous signature that the two beams are energy-time entangled. Thus, it can be used as a criterion for the test of the quantum nonlocal feature for continuous variable energy-time entanglement. In the following discussion, we will give our experimental results in a normalized form of Eq. (2), i.e.,

$$W = \frac{\langle(\Delta\tau')^2\rangle\langle(\Delta\tau)^2\rangle}{(\langle(\Delta\tau)^2\rangle)^2 + (2\beta l)^2} \geq 1 \qquad (3)$$

As discussed in Ref. [25], the jitter of the single-photon detectors (~tens of picoseconds), which is much greater than the biphoton correlation time, must be considered in practical experiment. Thus $\langle(\Delta\tau)^2\rangle$ in Eq. (3) should be replaced by $\langle(\Delta\tau)^2\rangle_{obs} = \langle(\Delta\tau)^2\rangle_{source} + \langle(\Delta\tau)^2\rangle_{jitter}$. When $\langle(\Delta\tau)^2\rangle_{jitter} \gg \langle(\Delta\tau)^2\rangle_{source}$, $\langle(\Delta\tau)^2\rangle_{jitter}$ becomes the dominant contribution to the observed time difference variance. Therefore, in order to violate the inequality, there are two ways that can be considered: one is to reduce the jitter time of the single photon detector, and the other is to increase the magnitude of dispersion. In our experiment, we realized the goal by using self-developed SNSPDs with a FWHM jitter as low as 37 ps and introducing a large amount of dispersion with a 62-km-long SMF.

Note that the shape of $G^{(2)}$ cannot be generally directly measured instead of the coincidence counting rate within a certain time window [41]. When the width of $G^{(2)}$ is much greater than the time window, the shape of $G^{(2)}$ can be observed by the distributions of coincidence counting from a TAC-MCA system [41] or TCSPC [27]. In our experiment, we will reconstruct the shape of $G^{(2)}$ using the the distributions of coincidence counting from the cross-correlation algorithm on time sequences tagged by ETs. Additionally, consider that only one dispersive media is connected into the channel, i.e., only the SMF in the signal arm or the DCF in the idler arm, then the FWHM of $G^{(2)}$ in the far-field regime can be approximated as $\Delta t_{s(i)} = \eta k''_{s(i)} l_{1(2)}$ with $\eta = \sqrt{2\ln 2/\gamma D^2 L^2}$. For a fixed biphoton source and dispersion coefficient of $k''_s$ or $k''_i$, $\Delta t_{s(i)}$ increases linearly with the length of SMF or DCF. Therefore, one can check whether there is a change of bandwidth of biphotons induced by the propagation channel loss by measuring the linear dependence of $\Delta t_{s(i)}$ on $l_{1(2)}$.

## III. EXPERIMENTAL RESULTS

### A. Experimental setup

The experimental setup is shown as Fig. 1. The energy-time entangled signal and idler photons are generated via SPDC within a piece of type-II PPKTP crystal (Raicol Ltd., $L$=1cm, a poling period of 46.146μm and $D$=2.96ps/cm) pumped by a 780 nm laser which is frequency doubled from a 1560 nm CW fiber laser (NKT photonics, Koheras BoostiK C15). Detailed information about the frequency doubling cavity can be found in Ref. [34]. The spectra of the signal and idler photons are centered at 1560.23 and 1560.04 nm while their 3-dB bandwidths being both 3.22 nm. After filtering out the residual pump, the output orthogonally polarized biphotons were coupled into the fiber polarization beam splitters (FPBS). With the help of a half-wave-plate (HWP) before the FPBS, the signal and idler photons were spatially separated for subsequent fiber transmission. Afterwards, the signal photons travelled through a SMF with $k''_s \sim -2.26 \times 10^{-26} s^2/m$ to site A and detected by detector $D_1$. The idler photons travelled to site B through a DCF with $k''_i \sim 1.95 \times 10^{-25} s^2/m$ and detected by detector $D_2$. $D_1$ and $D_2$ are SNSPDs with an efficiency of 50%. The arrival times of the signal and idler photons to the detectors were recorded independently by ET A and ET B (Eventech Ltd, A033-ET) as time tag sequences $\{t_A^{(j)}\}$ and $\{t_B^{(j)}\}$, which were transmitted through classical communication channel to the same terminal for data processing. In our experiment, the input time tag rate of each input was set around 12 kHz and a data acquisition time of 5s was applied. Thus both sequences for the signal and idler are about 60000. By applying cross-correlation algorithm on the acquired time sequences [36], the $G^{(2)}$ can be then constructed.

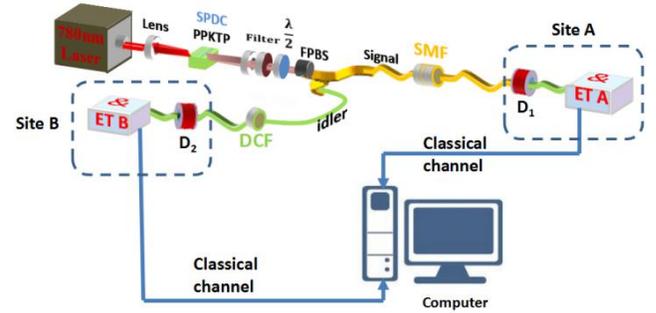

FIG. 1. (Color online) The experimental setup.

### B. Results

To determine the jitter contribution from the SNSPDs and the ETs, we firstly measured the coincidence distribution $G^{(2)}$ between the biphotons before fiber propagations. As shown in Fig. 2 (a), a FWHM width of about 37.6 ps was obtained. In order to verify the validity of our nonlocal coincidence detection method, we also measured directly the coincidence distributions with an integrated coincidence device PicoHarp 300, as shown in the inset of Fig. 2 (a). The very good agreement between the two results shows that the nonlocal detection method is highly suitable for the coincidence distribution measurement.

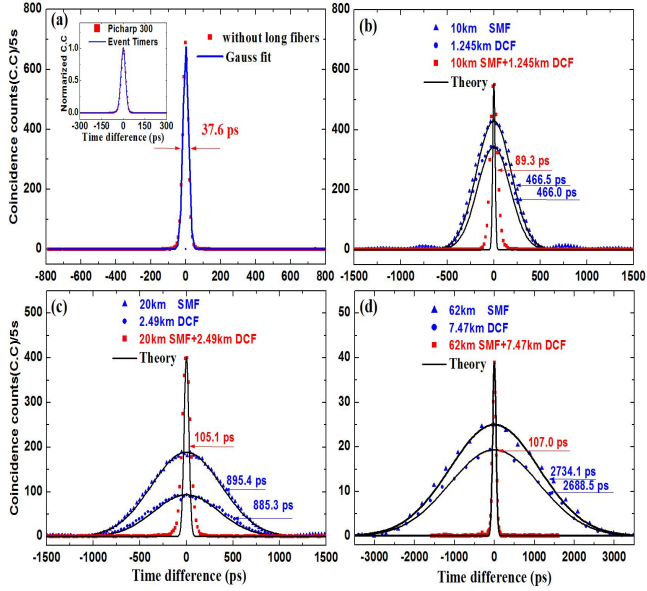

FIG. 2. (Color online) The reconstructed coincidence distributions from two ETs for (a) without long fibers, (b)-(c) with only SMF in the signal arm (blue triangles) or DCF in the idler arm (blue circles), both SMF and DCF connected (red squares). The lenghs of SMF/DCF were chosen as (b)10km/1.245km, (c)20km/2.49km, and (d) 62km/7.47km, respectively. The solid lines in (b)-(d) correspond to the theoretical results, while that in (a) is the Gauss fit. The inset in (a) gives the comparison of the normalized coincidence distributions obtained by Picoharp 300 and ETs, respectively.

Fig. 2 (b)-(d) show the results of coincidence counts with only SMF ($l_1$ correspondingly chosen as 10km, 20km and 62km) in the signal arm (blue triangles) or DCF ($l_2$ correspondingly chosen as 1.245km, 2.49km and 7.47km) in the idler arm (blue circles), both SMF and DCF connected (red squares). It can be seen that $G^{(2)}$ are broadened due to the fiber dispersion in the single arm. By comparing the broadening amounts caused by SMF and DCF for all three cases, $|k_s''l_1| \approx |k_i''l_2|$ is roughly satisfied. The FWHM of $G^{(2)}(\Delta t)$ (red squares) in the Fig. 2 (b)-(d) were narrowed to 89.3ps, 105.14ps and 107.0ps due to the NDC effect using both SMF and DCF in two arms (i.e., $k_s''l_2 + k_i''l_1 \approx 0$). The observed reduction of the coincidence counts from Fig. 2 (a)-(d) can be attributed to the channel loss as the length of the fiber increases. The corresponding theoretical simulations are also given by solid lines in Fig. 2 (b)-(d), which are in good agreement with the experimental results. Therefore, it can be concluded that the nonlocal coincidence measurement as well as the NDC have been successfully achieved.

Now we turn to whether the inequality given by Eq. (3) can be violated in our experiment. The inequality can not be violated due to insufficient dispersion in Fig. 2 (b)-(c). To violate the inequality, we introduce a large amount of dispersion with a 62-km-long SMF in Fig. 2 (d). From Fig. 2 (a), we achieved the time difference variance before dispersive propagation is $\langle(\Delta\tau)^2\rangle_{obs} = (15.982 \pm 0.150\text{ps})^2$. From Fig. 2 (d), we achieved the time difference variance after dispersive propagation through both 62km SMF and 7.47km DCF is $\langle(\Delta\tau')^2\rangle_{obs} = (45.676 \pm 4.565\text{ps})^2$. Using the average magnitude of the applied dispersions on both channels $2\beta l = 1428.92\text{ps}^2$, we obtained a value of $W = 0.253 \pm 0.052$. Therefore, the inequality given by Eq. (3) is obviously violated by about 14 standard deviations. Analogous to the Bell test for discrete variables, such violation indicates deterministically the nonlocal feature of the energy-time entanglement after propagating through 62-km-long fibers.

## IV. DISCUSSION

In our experiment, we utilized two individual ETs to tag the arrival times of the signal and idler photons. Since the time correlation between the propagated signal and idler photons is achieved via separate event timing systems instead of a local coincidence hardware (e.g. Picoharp 300), such detection method is eventually nonlocal and is desired for the real-field quantum nonlocality test. As mentioned above, the amount of time sequences is limited by the ETs. By manipulating the pump power of the SPDC and adding appropriate loss in the idler arm, we fixed all the detected single count rates as about 60000 per 5 seconds.

According to Eq. (2), the width of $G^{(2)}$ with perfect NDC should be equivalent to the coherence width of the generated biphotons, which is about 2.96ps in our case [34]. However, due to the inherent time jitter of single-photon detectors, the minimum width of $G^{(2)}$ that can be achieved is limited to about dozens of picoseconds (37.6 ps in our experiment). By looking at the results shown in Fig. 2 (d) and Fig. 3, both the measured and simulated $G^{(2)}$ indicate the incompleteness of NDC, that is, there is residual uncompensated dispersion resulting in broadening of the $G^{(2)}$ wave packet. To achieve perfect dispersion cancellation, a dispersion-adjustable component such as chirped fiber Bragg gratings [42] should be desired to overcome under-compensation or over-compensation for satisfying the compensation condition as much as possible.

Another issue to be considered is that the reduction of bandwidth of biphotons may also lead to the broadening of $G^{(2)}$ [27]. So, is it due to the dispersion of fibers or the reduction of bandwidth of biphotons? In order to answer this question, we give an experimental result of $\Delta t$ versus the length of SMF and DCF in single arm as shown in Fig. (3). We can see that $\Delta t$ increase linearly with the length of SMF and DCF, which agree well with the theoretical prediction. From the linear fits, we obtained a slope of 42.96ps/km and 359.63ps/km, and derived the dispersion coefficient being $k_s'' \simeq -2.37 \times 10^{-26} \mathrm{s}^2/\mathrm{m}$ and $k_i'' \simeq 1.99 \times 10^{-25} \mathrm{s}^2/\mathrm{m}$, which are also in good agreement with the experimental parameters. Thus, we give a proof that the broadening of $G^{(2)}$ in single arm is only attributed to the length of optical fibers instead of the reduction of bandwidth of biphotons in spite of channel loss.

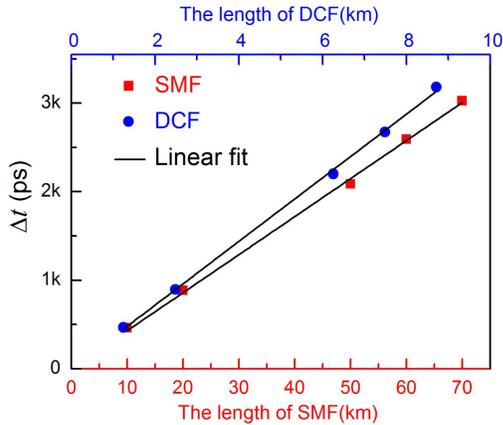

FIG. 3. (Color online) The reconstructed FWHM of $G^{(2)}$ ($\Delta t$) versus the length of SMF and DCF in single arm.

Note that when the signal or idler photons propagate through a dispersive channel, random noise associated with the dispersion add to the registered time sequences, and decreases the photon pair collection efficiency. However, benefited from the NDC effect, the narrowed $G^{(2)}$ was observed. Therefore, the NDC enables remaining tightly bunched of arrival times of the signal and idlers after long-distance transmission, and thus recovering the efficient signal from the noise and improving the signal-to-noise ratio.

Our NDC is implemented at telecom wavelength, which is robust in transmission through both fiber and free-space links, and is compatible with wavelength-division multiplexing. Although our result is merely a proof-of-principle experiment on the nonlocality test, it demonstrates the feasibility for the strict test of nonlocal feature of continuous variable entanglement with the energy-time entangled photons over long distance. Furthermore, the NDC can also be applied to enhance the stability of quantum time transfer [43].

## V. CONCLUSION

In summary, we reported a picosecond-level NDC and demonstrated a violation of Wasak's inequality for determining the nonlocal feature of energy-time entanglement with a nonlocal detection method over 60km fibers at telecom wavelength. According to the violation criterion given by Eq. (3), we obtained a value of $W = 0.253 \pm 0.052 < 1$, which violates the inequality obviously by about 14 standard deviations. We believe our work is a significant step toward the future strict test of nonlocality for continuous variable entanglement after a long-distance fiber or free space transmission.

## ACKNOWLEDGMENTS

This work was supported by the National Natural Science Foundation of China (Grant Nos. 61875205, 91836301, 11273024, 91636101, 61025023, 61801458, and 11504292), the Research Equipment Development Project of Chinese Academy of Sciences (Project Name: Quantum Optimization Time Transfer Experiment System Based on Femtosecond Optical Frequency Combs), the National Youth Talent Support Program of China (Grant No.[2013] 33), the National Key R&D Program of China (2017YFA0304000) and the Frontier Science Key Research Project of Chinese Academy of Sciences (Grant Nos. QYZDB-SSW-SLH007 and QYZDB-SSW-JSC013), Natural Science Basic Research Plan in Shaanxi Province of China (Grant No. 2019JM-346), and Outstanding Youth Science Fund of Xi'an University of Science and Technology (Grant No. 2019YQ2-13).